\documentclass[fleqn,usenatbib]{mnras}
\usepackage{newtxtext,newtxmath}

\usepackage[T1]{fontenc}
\usepackage{longtable}

\usepackage{graphicx}
\usepackage{bm}
\usepackage{natbib}
\usepackage{booktabs}
\usepackage{color}
\usepackage{units}
\usepackage{xcolor}

\usepackage[normalem]{ulem}

\newcommand{\program}[1]{\textsc{#1}}
\newcommand{\citeg}[1]{\citep[e.g.,][]{#1}}
\newcommand{\be}{\begin{equation}}
\newcommand{\ee}{\end{equation}}

\newcommand{\ciera}{Center for Interdisciplinary Exploration and Research in Astrophysics (CIERA), Northwestern University, 1800 Sherman Ave., Evanston, IL 60201, USA}

\title[Type II surrogates]{Surrogate models for lightcurves and photosphere properties of Type II supernovae}

\author[N.~Sarin et al.]{Nikhil Sarin$^{1}$\thanks{E-mail: nsarin.astro@gmail.com},
Takashi J. Moriya$^{2,3,4}$,
Avinash Singh$^{5,6}$,
Anjasha Gangopadhyay$^{5}$,
K-Ryan Hinds$^{7}$,
\newauthor
Steve Schulze$^{8}$,
Conor M. B. Omand$^{7}$,
and Kaustav K. Das$^{9}$ 
\\
$^{1}$Oskar Klein Centre for Cosmoparticle Physics, Department of Physics, Stockholm University, AlbaNova, Stockholm SE-106 91, Sweden\\
$^{2}$National Astronomical Observatory of Japan, National Institutes of Natural Sciences, 2-21-1 Osawa, Mitaka, Tokyo 181-8588, Japan \\
$^{3}$Graduate Institute for Advanced Studies, SOKENDAI, 2-21-1 Osawa, Mitaka, Tokyo 181-8588, Japan \\
$^{4}$School of Physics and Astronomy, Monash University, Clayton, VIC 3800, Australia \\
$^{5}$The Oskar Klein Centre, Department of Astronomy, Stockholm University, Albanova, 106 91 Stockholm, Sweden \\
$^{6}$Hiroshima Astrophysical Science Center, Hiroshima University, Higashi-Hiroshima, Hiroshima 739-8526, Japan \\
$^{7}$Astrophysics Research Institute, Liverpool John Moores University, Liverpool Science Park IC2, 146 Brownlow Hill, Liverpool, UK, L3 5RZ\\
$^{8}$\ciera\\
$^{9}$Cahill Center for Astrophysics, California Institute of Technology, MC 249-17, 
1200 E California Boulevard, Pasadena, CA, 91125, USA
}

\date{Accepted XXX. Received YYY; in original form ZZZ}

\pubyear{\the\year{}}

\begin{document}
\label{firstpage}
\pagerange{\pageref{firstpage}--\pageref{lastpage}}
\maketitle

\begin{abstract}
Inferences on the properties Type II supernovae (SNe) can provide significant insights into the lives and deaths of the astrophysical population of massive stars and potentially provide measurements of luminosity distance, independent of the distance ladder. Here, we introduce surrogate models for the photospheric properties and lightcurves of Type II SNe trained on a large grid of simulations from the radiation hydrodynamics code, {\sc stella}. The trained model can accurately and efficiently ($\sim 30$ms) predict the lightcurves and properties of Type II SNe within a large parameter space of progenitor ($10-18 M_{\odot}$ at ZAMS) and nickel masses ($0.001-0.3M_{\odot}$), progenitor mass-loss rate ($10^{-5}-10^{-1}~M_{\odot}$yr$^{-1}$), CSM radius ($1-10\times10^{14}$cm), and SN explosion energies ($0.5-5 \times 10^{51}$erg). We validate this model through inference on lightcurves and photosphere properties drawn directly from the original {\sc stella} simulations not included in training. In particular, for a synthetic Type II SNe observed within the 10-year LSST survey, we find we can measure the progenitor and nickel masses with $\approx 9\%$ and $\approx 25\%$ precision, respectively, when fitting the photometric data while accounting for the uncertainty in the surrogate model itself. Meanwhile, from real observations of SN~2004et, SN~2012aw, and SN~2017gmr we infer a progenitor ZAMS mass of $12.15_{-1.06}^{+1.03} M_{\odot}$, $10.61_{-0.32}^{+0.37} M_{\odot}$, $10.4 \pm 0.3 M_{\odot}$, respectively. We discuss systematic uncertainties from our surrogate modelling approach and likelihood approaches to account for these uncertainties. We further discuss future extensions to the model to enable stronger constraints on properties of Type II SNe and their progenitors, and applications of our surrogate modelling approach to other transients. 
\end{abstract}

\begin{keywords}
supernovae: general -- stars: supergiants -- methods: statistical 
\end{keywords}

\section{Introduction}\label{sec:intro}
Type II supernovae (SNe) are explosions of massive stars that retain significant hydrogen envelopes, and they represent the most common class of core-collapse SNe observed in the local Universe~\citeg{Hinds2025}. Their light curves and spectral evolution encode valuable information about their progenitor stars, explosion mechanisms, and surrounding circumstellar matter (CSM)~\citep[see e.g.,][for a review]{Janka2007, Jerkstrand2025}. With the advent of large-scale optical transient surveys like the Zwicky Transient Facility \citep[ZTF;][]{Bellm2019} and the upcoming Vera C. Rubin Observatory's Legacy Survey of Space and Time (LSST)~\citep{Ivezic2019}, the discovery rate of Type II SNe is increasing dramatically, promising unprecedented sample sizes for statistical studies. For example, recently~\citet{Das2025} published a sample of 330 Type IIP (so called due to the plateau in their lightcurves) SNe from a systematic volume-limited (out to 150 Mpc) survey with ZTF, while \citet{Hinds2025} compiled a larger sample of $639$ Type II SNe in a magnitude-limited ZTF survey. These numbers are predicted to grow by more than an order of magnitude with LSST discovering events out to a larger volume~\citep[e.g.,][]{Ivezic2019}. 

Lightcurve and spectral modelling of past Type II SNe has yielded insights into the explosion energy, nucleosynthetic yields and progenitor properties of massive stars~\citep{Forster2018, Davies2018, Goldberg2019, Martinez2022, Subrayan2023, Silva-Farfan2024, Hinds2025}. These properties have been used to elucidate the explosion mechanism~\citep[e.g.,][]{Burrows2021}. As the dataset continues to grow through time-domain surveys, we are increasingly in need of computationally efficient models to extract physical parameters from often sparsely sampled photometric observations. While detailed numerical simulations provide the most physically grounded models~\citep[e.g.,][]{Bersten2011, Dessart2013, Moriya2023}, their significant computational expense prohibits direct application for large-scale inference for hundreds or thousands of events. Meanwhile, simpler analytical or semi-analytical models~\citep{Popov1993, Nagy2016}, while faster, often rely on simplifying assumptions, such as uniform density profiles, spherical symmetry, constant opacity, and simplified energy deposition from radioactive decay. These simplifications limit their ability to adequately capture the complexity revealed by detailed simulations and probed by detailed observations. This is especially true in cases where the lightcurve may be powered by CSM interaction or behave differently due to different assumptions on ${}^{56}$Ni mixing~\citep[e.g.,][]{Forster2018, Bruch2021}. 

\citet{Moriya2023} recently published a large grid of synthetic Type II SN models based on red supergiant progenitors (RSG) from~\citet{Sukhbold2016}, including properties such as the photospheric temperatures and radius, bolometric luminosities, as well as spectrum computed through the radiation hydrodynamics code~{\sc stella}~\citep{Blinnikov1998, Blinnikov2000, Blinnikov2006}. The grid from \citet{Moriya2023}, encompassing 228,016 models, systematically explores a wide parameter space, including progenitor zero age main sequence (ZAMS) mass, explosion energy, $^{56}$Ni mass, and various wind-like CSM properties (mass-loss rate, extent, structure). Comparisons with this grid have already facilitated a wide variety of analyses into the properties of Type II SNe~\citep[e.g.,][]{Moriya24,Chen2025,Hinds2025}. However, while this grid forms a crucial basis for systematic studies, direct interpolation within such a high-dimensional grid for parameter inference can still be challenging, preventing the use of these models to directly fit lightcurves and extracting posteriors on explosion and progenitor properties. 

\begin{figure*}
    \centering
    \includegraphics[width=0.95\textwidth]{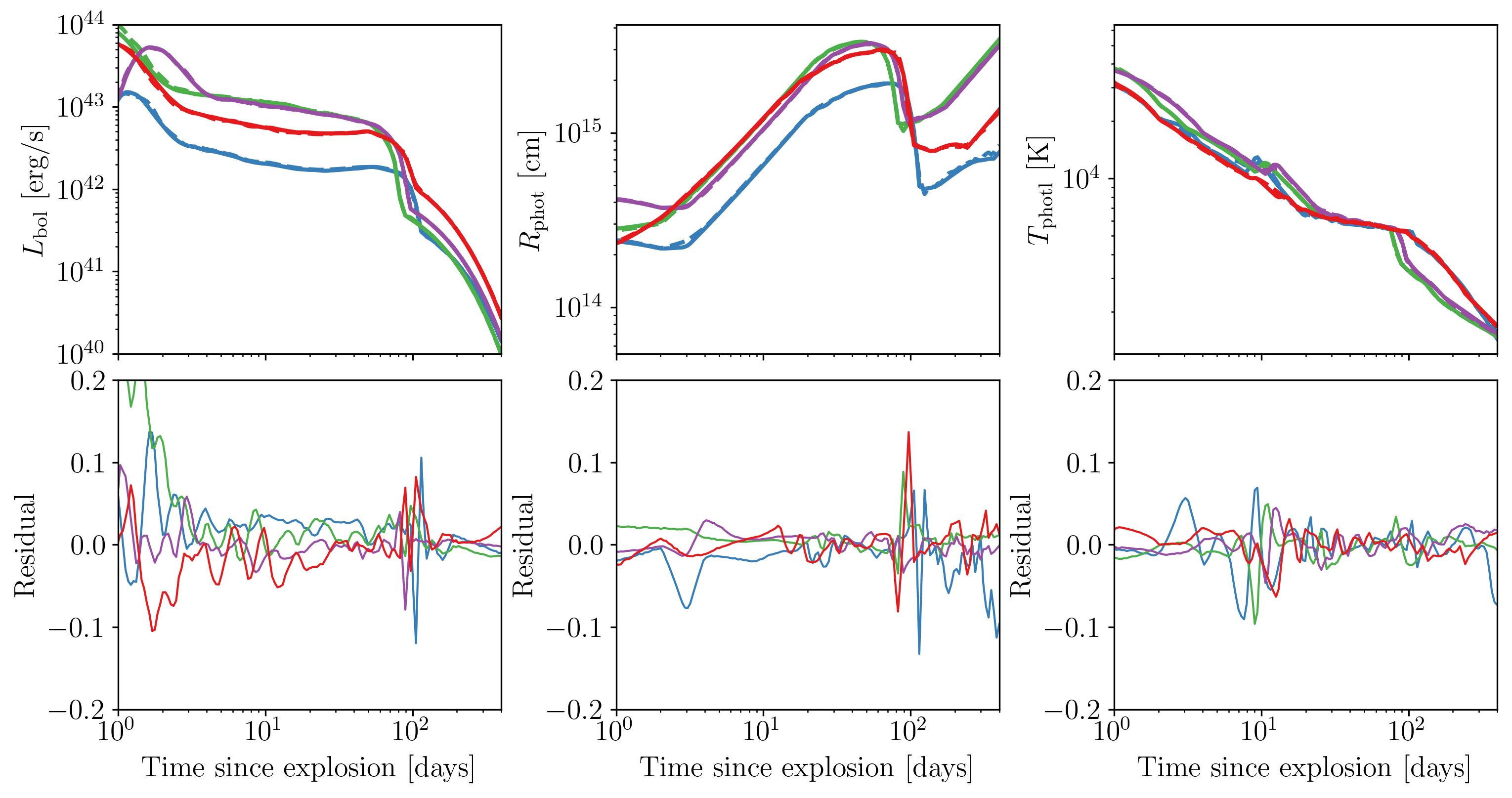}
    \caption{Bolometric luminosity and photosphere temperature and radius for random samples from our training data (solid curves) alongside the surrogate model predictions (dashed curves), while the bottom panels show the normalised residuals on each quantity.}
    \label{fig:validation}
\end{figure*}

Here, we introduce a new surrogate models for the spectra, lightcurves and photosphere properties from this grid, that can be used to efficiently and accurately interpolate the outputs from the numerical simulations. In particular, in Sec.~\ref{sec:MLemulator}, we describe the design, training process, and performance of our surrogate models. In Sec.~\ref{sec:validation}, we validate our surrogate models by performing inference on bolometric luminosity lightcurve and photosphere properties against the original {\sc stella} simulations, as well as for synthetic photometric observations with the Vera Rubin Observatory. We then demonstrate the utility of the model by inferring the physical parameters of a few Type II SNe by fitting their photometric observations in Sec.~\ref{sec:application}. 
Finally, we discuss future extensions to our models, systematic uncertainties associated with the surrogate modelling approach, and conclude in Sec.~\ref{sec:conclusion}.

\section{Surrogate model}\label{sec:MLemulator}
\subsection{Training grid assumptions}
As mentioned above,~\citet{Moriya2023} provide us a total of $228,015$ different simulations\footnote{We note that one simulation from the original 228,016 simulations is missing bolometric luminosity data and could not be run again.}. We briefly describe the primary ingredients, modelling assumptions, and limitations below, but refer the interested reader to \citet{Moriya2023, Moriya24} for further details. All parameters of the model, including their description and ranges are summarised in Table~\ref{tab:parameters}.

The model adopts red supergiant progenitors from \citet{Sukhbold2016} and attach a CSM of the form 
\begin{equation}
\rho_{\rm CSM} (r) = \frac{\dot{M}}{4\pi v_{\rm wind}(r)}r^{-2},
\end{equation}
where the velocity, $v \propto (1 - R_{0}/r)^{\beta}$, where $R_0$ is the progenitor radius and $\beta$ sets the steepness of the density profile. This CSM structure extends out to $R_{\rm CSM}$, with the maximum value of $R_{\rm CSM} = 10^{15}$cm motivated by observations~\citep{Moriya2023}. The model then further assumes some explosion energy $E_{\rm SN}$ and a mass of $^{56}$Ni that is uniformly mixed up to the half mass of the hydrogen-rich envelope. Lightcurves and photosphere properties are then generated using~{\sc stella}~\citep{Blinnikov1998, Blinnikov2000, Blinnikov2006}.

In general, this setup is well suited to capture the broadband properties of Type II SNe. However, there are a number of limitations with this setup which could affect our conclusions when applied to real data. As discussed in \citet{Moriya24}, the choice of progenitor model and parameterisation has an impact on the constraints on the progenitor mass due to the strong correlation of this parameter with $R_{0}$, which is implicitly fixed by the choice of progenitor model. Moreover, as the radius out to which $^{56}$Ni is distributed is fixed across all simulations, we can not capture any discrepancies in the lightcurve due to this effect alone. Furthermore, at late-times as the supernova becomes nebular, the local-thermodynamic equilibrium (LTE) assumption built into \program{stella} breaks down, making it unsuitable for estimating the properties of Type II SNe at $\gtrsim 100$ days post explosion.

\subsection{Training data and surrogate model design}\label{subsec:trainingdataandmodel}
Each simulation from \citet{Moriya2023} contains data for the bolometric luminosity, photosphere temperature, photosphere radius, and spectral energy distribution (SED) at $100$ wavelengths from $1-50000$~\AA{} at multiple epochs in time. 
Ideally, a single surrogate model trained directly on the SED can be used to reconstruct photometry and bolometric luminosities. Moreover, it can capture the effects of any emission or absorption lines in the spectrum, enabling fitting directly to spectroscopic data. 
In practice, the SEDs in our training data do not include any spectral features, they are also not sufficiently densely sampled to enable detailed inferences against high-resolution spectra. Therefore, we build multiple independent surrogates for the bolometric luminosity, photosphere properties (temperature and radius), and for the spectrum itself which can be used to generate photometry. This approach provides flexibility enabling us to perform inference on different types of data (or combine multiple diagnostics) which can be used to further test models. 
We note that the photosphere temperature and radius could also be used to generate photometry assuming a blackbody SED, but Type II SNe are not well approximated by a blackbody at late phases or at ultra-violet (UV) wavelengths, so we do not recommend this approach. 

\subsubsection{Luminosity and photosphere property surrogates}
Our first step towards building our surrogate models was to prepare the training data, as each simulation output is on different time samples, we processed all the different outputs (bolometric luminosity, photosphere temperature, and photosphere radius) for all simulations onto a consistent time array from 0.1 to 400 days (in rest frame time). We do this through linear interpolation as the simulation outputs are densely sampled and smooth functions. 

We randomly split the $228,015$ simulations into a testing set ($30,000$ simulations) and training set ($198,015$ simulations). The training set is further divided into training ($158,412$ simulations, i.e., $80\%$ of the training set) and validation ($39,603$ simulations, i.e., $20\%$ of the training set) subsets for model development and hyperparameter tuning.
We also transform the input parameters and outputs onto a unit Gaussian to aid training. The validation simulations are used during each epoch of training to validate the model that is trained on the training set. The testing data is used after training is finished to test the final trained model.

\begin{table}
    \centering
    \begin{tabular}{|l|p{3cm}|l|}
        \hline
        \textbf{Parameter} & \textbf{Description} & \textbf{Range} \\
        \hline
        $M_{\rm ZAMS}$ [$M_{\odot}$] & Initial ZAMS mass of the progenitor & 10 - 18 \\
        \hline
        ${}^{56}$Ni [$M_{\odot}$]& Mass of nickel-56 & 0.001-0.3 \\
        \hline
        $\dot{M}$ [$M_{\odot}$yr$^{-1}$]& Mass loss rate of the progenitor & $10^{-5}-10^{-1}$ \\
        \hline
        $\beta$ & Steepness of the CSM density profile & 0.5-5 \\
        \hline
        $R_{\rm csm}$ [$10^{14}$cm]& Radius of the circumstellar material & 1-10 \\
        \hline
        $E_{\rm sn}$ [$10^{51}$erg]& Energy of the supernova explosion & 0.5-5 \\
        \hline
    \end{tabular}
    \caption{Input parameters, descriptions, and the ranges of the surrogate model}
    \label{tab:parameters}
\end{table}
As all simulations are on discrete points, e.g., $10, 12, 14, 16, 18\ M_{\odot}$ for the ZAMS mass, we also augment our training data by adding Gaussian noise ($\sigma = 0.01$ in normalized parameter space) to both inputs and outputs for each simulation, effectively creating 5 additional samples per original simulation. This augmentation increases our training set size to $\approx 800,000$ samples. We then employ a feed forward neural network via {\sc tensorflow}~\citep{Abadi2016} with a combination of up to 5 layers with a hyperbolic tangent activation function and a combination of mean absolute and mean squared error loss functions. We train with a batch size of 5000 (to fully leverage GPUs) and for up to 400 epochs (with early stopping, in the case the model does not improve for more than 10 epochs). Our total training time is $\approx 10$ mins on a {\sc M1 pro} Macbook GPU, with the output being three different trained models which independently predict the bolometric luminosity and photosphere temperature and radius. 
The augmentation has a significant impact on the accuracy of our trained model on the unseen testing set, improving loss by $\gtrsim 5\%$, while also enabling the model to generalize more reliably between discrete grid points.
\
\subsubsection{Spectrum surrogate}
To create a surrogate model for the spectrum we use a combination of an autoencoder, a principle-component analysis, and a feed-forward neural network. We first preprocess our spectrum data onto uniform arrays in time (geometrically from $0.1-400$ days) and wavelength(geometrically from $500-49500$~\AA{} using a cubic interpolator. We have verified that this faithfully captures the true spectrum across a range of simulations. We also normalise the flux data such that all values are between $0-1$ and augment our data as described above. The autoencoder encodes our uniform high-dimensional flux data into a $64$ dimensional latent-space, while also training a decoder to invert this process. We further reduce the dimensionality by performing a principal component analysis to $32$ dimensions with minimal impact on our accuracy. We then train a feed-forward neural network (with similar design to our surrogates above) to learn the mappings from the $6$ physical input parameters onto the reduced $32$ dimensional latent space. The full pipeline takes $\approx 40$~mins to train with adapting batch sizes (to ensure more granular gradient updates towards the end of training) and a total of 100 epochs. Our surrogate model can then be used to generate the spectrum for any arbitrary combination of parameters (within the ranges specified in Table~\ref{tab:parameters}), we can then generate photometry by integrating the spectrum over the bandpass of any filter via {\sc Redback}~\citep{Sarin2024} utilising filter transmission curves through the Spanish Virtual Observatory~\citep{Rodrigo2020}. 

\begin{figure}
    \centering
    \includegraphics[width=\linewidth]{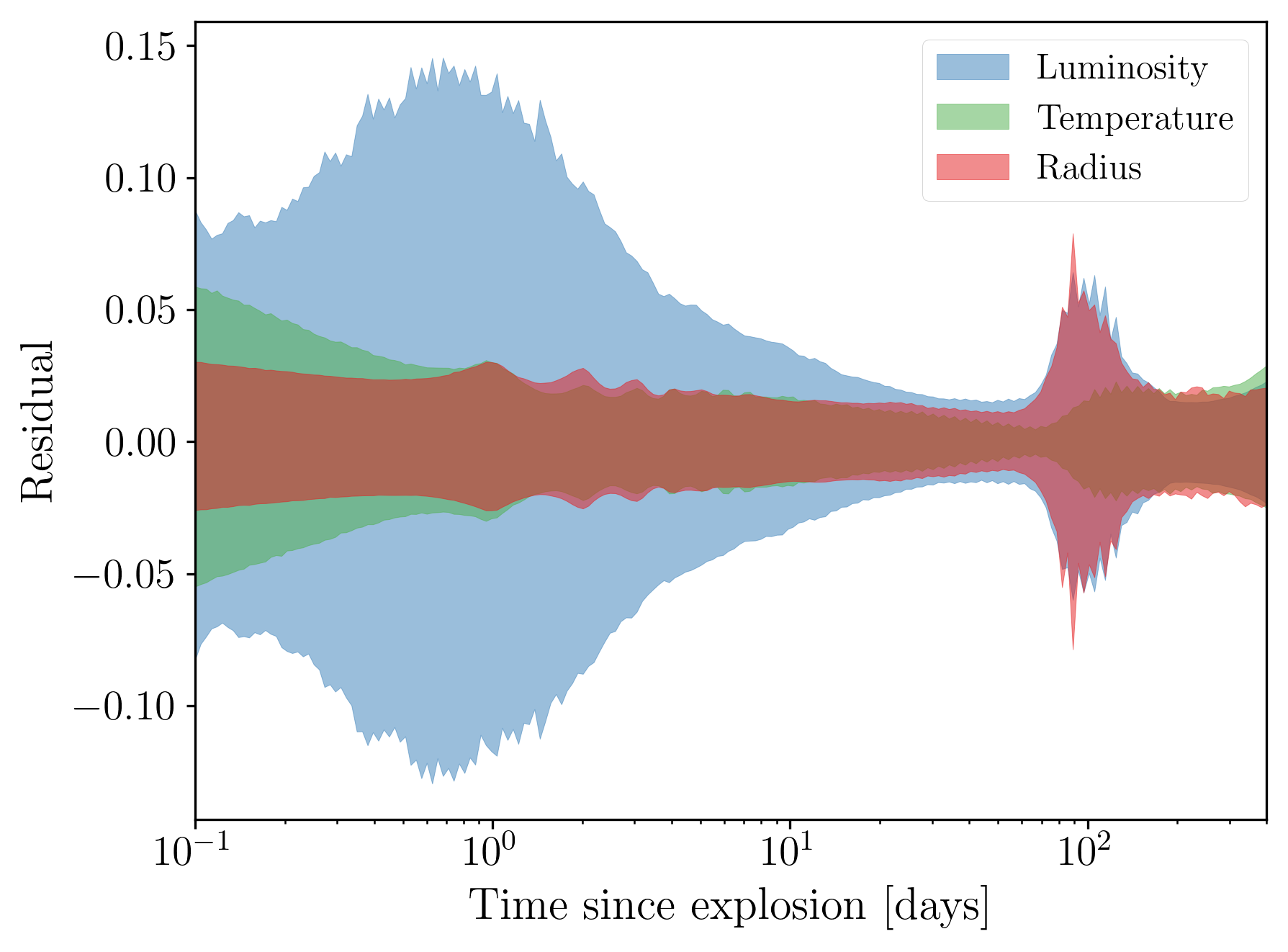}
    \caption{$1\sigma$ credible interval of the normalised residuals of our surrogate model predictions compared to the testing data.}
    \label{fig:residuals}
\end{figure}

\begin{figure*}
    \centering
    \includegraphics[width=0.95\textwidth]{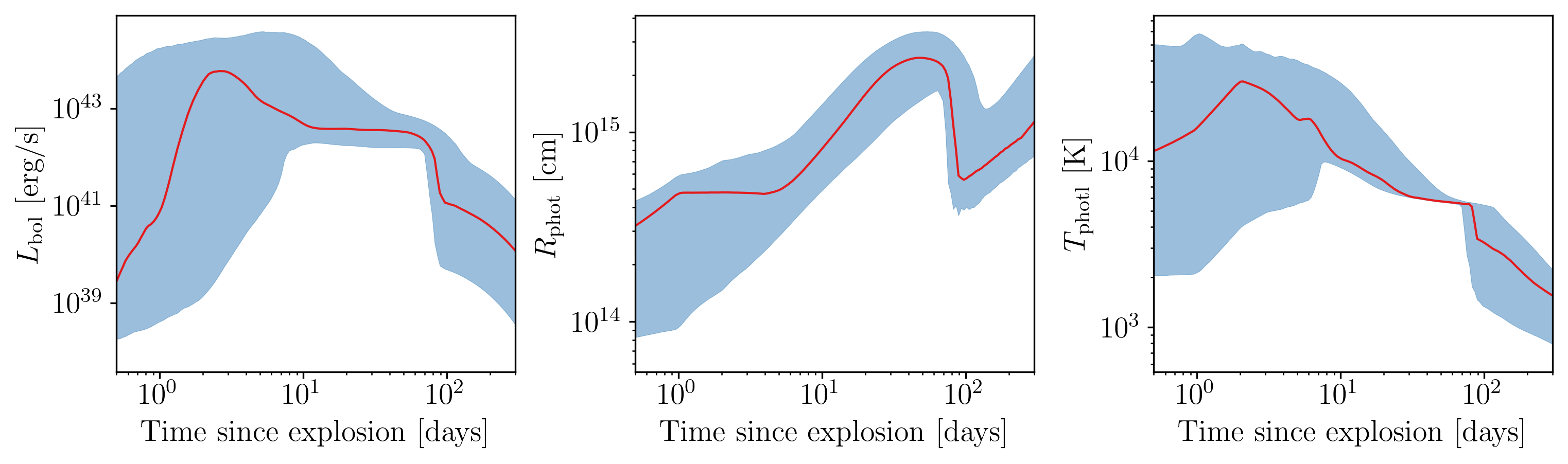}
    \caption{$90\%$ credible interval (shown by blue shaded region) of the bolometric luminosity, photosphere temperature and radius generated from $50000$ random samples our prior. We also show the predicted properties for a Type II SN with $M_{\rm ZAMS}=13~M_{\odot}$, ${}^{56}\rm{Ni}=0.02~M_{\odot}$, $\dot{M}=10^{-3.1}$~$M_{\odot}$yr$^{-1}$, $\beta=1.2$, $R_{\rm csm} = 5.5\times10^{14}$cm, and $E_{\rm sn}=2.1\times10^{51}$erg (red curves), a sample} not included in our training or simulated data. 
    \label{fig:priordraws}
\end{figure*}

\subsection{Performance}
\subsubsection{Luminosity and photosphere property surrogates}
To evaluate the performance of our trained models, we consider a range of metrics relative to an unseen training data of $30000$ simulations. Our first consideration is accuracy, across the training data, we see mean squared errors of $0.0056$, $0.06$, and $0.01$ for the luminosity, temperature, and radius surrogates, respectively, on the normalized values. 
Moreover, our models are also extremely fast, with each individual prediction taking $\approx 30$ms on a standard CPU. Furthermore, due to the neural network architecture's parallel processing capabilities, we can generate predictions for large batches simultaneously. For example, all outputs for the $30,000$ test simulations are generated within $\approx 30$ms, which can provide significant benefits for inference and population studies as we will discuss later.
In Fig.~\ref{fig:validation}, we plot a random set of bolometric luminosities, photosphere radius and temperature from our training data (solid curves) alongside the prediction of the surrogate model (dashed curves). The bottom panel indicates the normalised residual (which we explore in more detail below), which highlights that across the bulk of times, our normalized residuals are smaller than $5\%$, this is smaller than the numerical noise in the simulations itself, particularly at early times. 

We further investigate the residuals across our test data to see if there are any patterns of failure with our surrogate model. In particular, in Fig.~\ref{fig:residuals}, we show $1\sigma$ distribution on the normalised residuals as a function of time from the full sample of our $30000$ testing data. First, we draw attention to the luminosity residuals, which indicates a discrepancy between our surrogate model and the testing data of as large as $15\%$ at $t \sim 1$d, which could influence our results when performing inference if unaccounted for. Our residuals also highlight a peak at $\approx 100$days for all models. This broadly coincides with the rapid drop in radius and luminosity in many models around this epoch, and the residual is a product of this sharp change and the numerical noise in multiple simulations at these epochs. However, as this is only on the $\approx 5\%$ level, we consider this accurate for our purposes of inferences, as opposed to the larger error in the bolometric luminosity surrogate at early times ($t \lesssim 1$day). We note that while Fig.~\ref{fig:residuals} shows the $1\sigma$ credible interval, our median normalised residuals are significantly smaller and on the order of $10^{-3}$, highlighting strong consistency between the surrogate model predictions and the training data.

Our ultimate goal with a surrogate model is to efficiently draw new model predictions and perform inference on real data. To investigate how our surrogate model performs when interpolating, we randomly sample from a uniform prior on all parameters (uniform over the ranges specified in Table~\ref{tab:parameters}). We note that while the model could also be used to extrapolate, this is not reliable due to the feature scalings we performed when preprocessing the simulations. In Fig.~\ref{fig:priordraws}, the blue bands show the $90\%$ credible interval from $50,000$ random samples from our prior for the bolometric luminosity, photosphere radius, and photosphere temperature for times between $0.1$ to $400$ days. This immediately provides some useful diagnostics for observations and interpretation. For example, at $10$ days, the bolometric luminosity of a Type II SN (for our choice of prior) is between $3\times 10^{42}-2\times10^{43}$erg/s. Furthermore, as we are now not limited by the need to perform additional simulations, we can easily explore the effect of changing parameters or make predictions for parameters where there is no simulated data. For example, we show predictions (red curves) for a Type II SN with $M_{\rm ZAMS}=13~M_{\odot}$, ${}^{56}\rm{Ni}=0.02~M_{\odot}$, $\dot{M}=10^{-3.1}$~$M_{\odot}$yr$^{-1}$, $\beta=1.2$, $R_{\rm csm} = 5.5\times10^{14}$cm, and $E_{\rm sn}=2.1\times10^{51}$erg, a sample not in the original $228,015$ simulations.

\begin{figure}
    \centering
    \includegraphics[width=0.87\columnwidth]{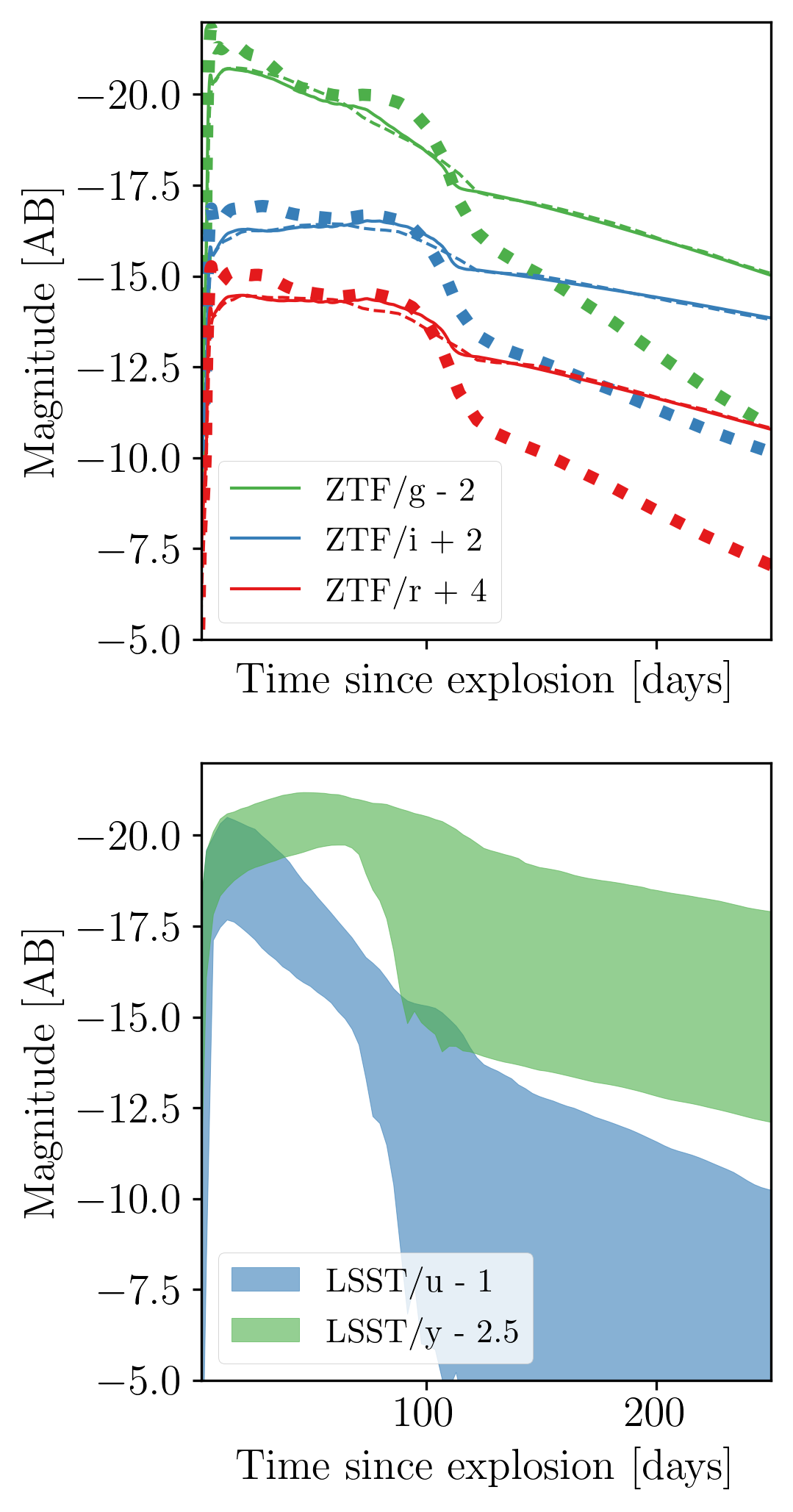}
    \caption{The top panel shows the true absolute magnitudes (solid curves) for a range of ZTF filters with our surrogate model prediction shown with dashed curves and the prediction assuming a blackbody SED (in squares). The bottom panel shows a $90\%$ credible interval (shown by blue shaded region) absolute magnitudes against time (in observer frame) for the lsst/u (blue) and lsst/y (green) bands drawn from $5000$ random samples from our prior.}
    \label{fig:absmag}
\end{figure}

\subsubsection{Spectrum surrogate}
Although properties such as the bolometric luminosity and photosphere properties like temperature and radius are useful, the bulk of current observations are from surveys with a few bandpasses. 
We therefore also want to predict photometry from our model. As we discussed in Sec.~\ref{subsec:trainingdataandmodel}, assuming a blackbody SED is problematic, particularly at late phases when the SN starts to become nebular or at ultra-violet wavelengths where line-blanketing strongly suppresses the flux compared to a blackbody. Here, we find our autoencoder-regressor approach to be sufficiently accurate and efficient. In particular, comparing with a testing set ($25\%$ of all data) kept out of training, we find our autoencoder achieves a reconstruction mean squared error of $0.001$, i.e., we are accurately representing the complex high-dimensional spectrum with just $64$ dimensions. Independently, the regression model achieves a mean squared error of $0.02$ indicating that a feed-forward neural network can accurately learn the mappings between the physical parameters and the latent space. Overall, the loss on the full pipeline is $\approx 5\%$ on the full spectrum. 

\begin{figure}
    \centering
    \includegraphics[width=0.95\columnwidth]{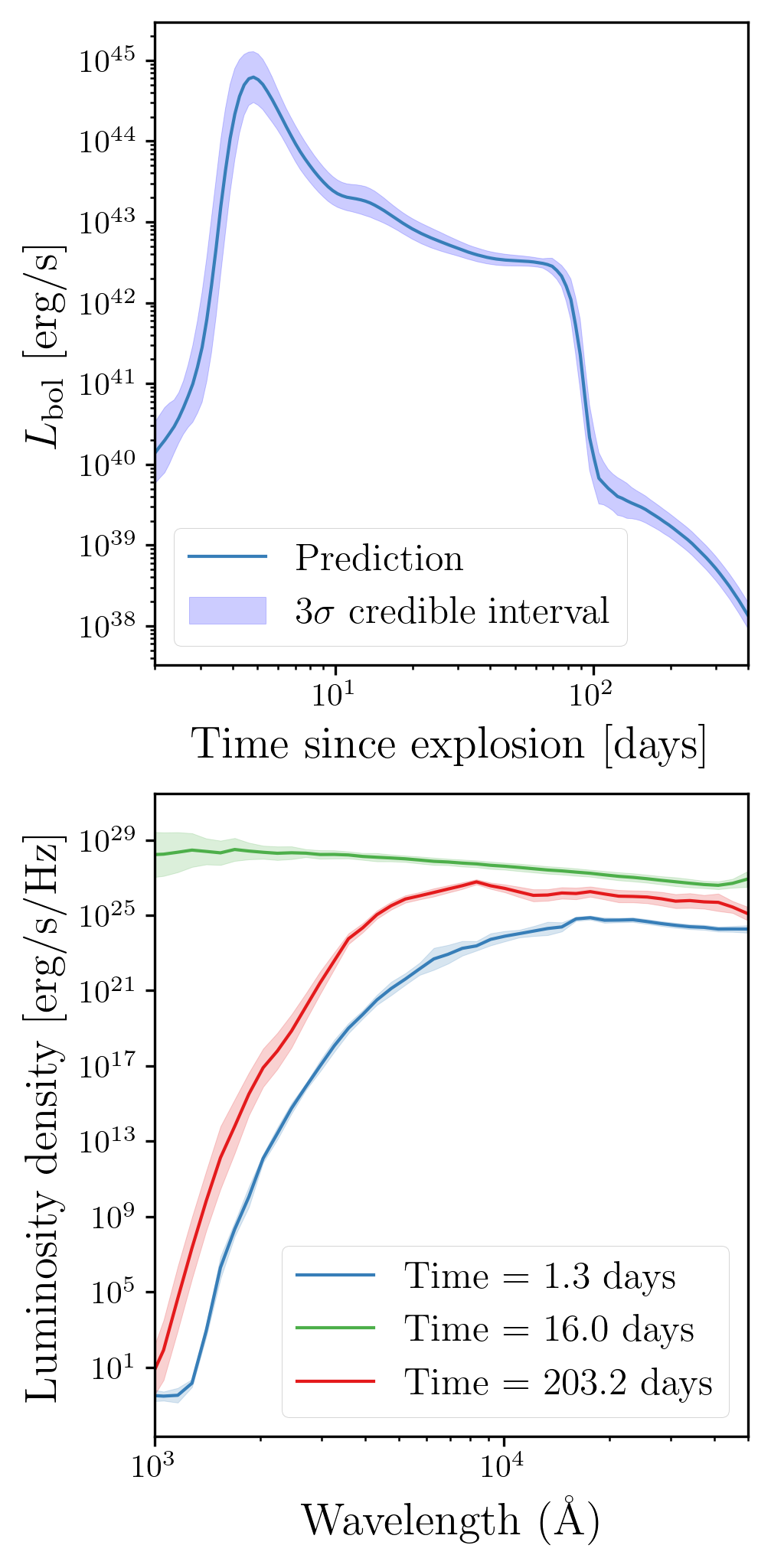}
    \caption{Example of a bolometric luminosity (top panel) and spectrum predictions (bottom panel) from our model alongside an estimate of the uncertainty estimated through Monte Carlo sampling ($3\sigma$ uncertainty shown by shaded bands).}
    \label{fig:uncertaintyestimates}
\end{figure}

\begin{figure*}
    \centering
    \includegraphics[width=0.95\textwidth]{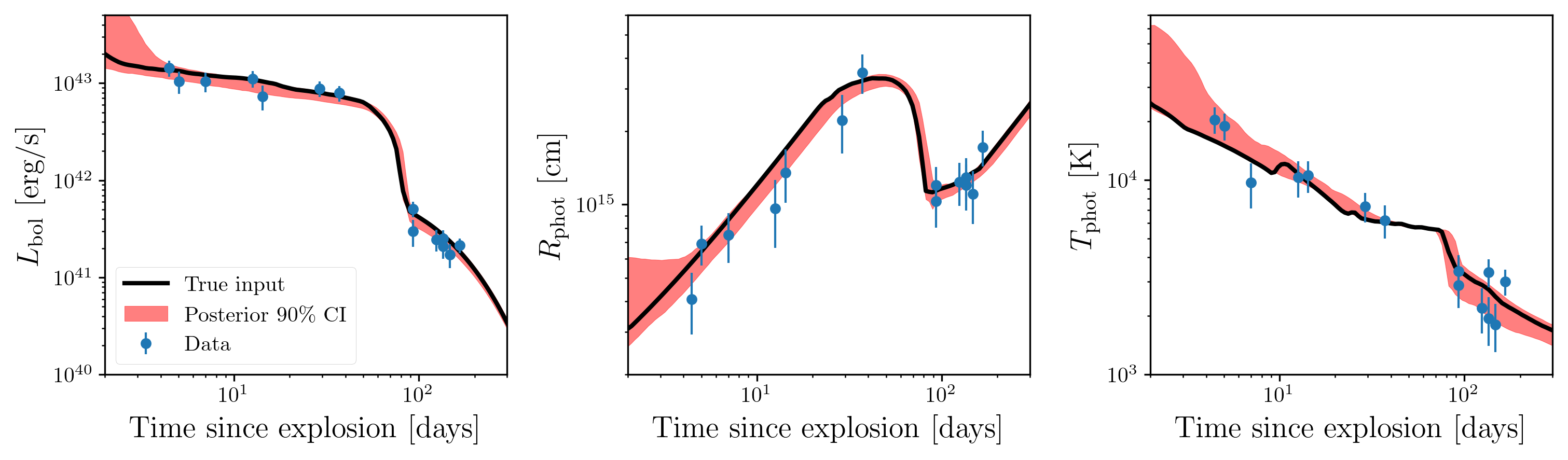}
    \caption{True input (in black) for the bolometric luminosity, photosphere radius, and photosphere temperature respectively from a {\sc stella} simulation and synthetic data (blue) obtained by adding Gaussian noise to the input. The red shaded bands represent the $90\%$ credible interval of the posterior prediction.}
    \label{fig:sim_allthree}
\end{figure*}

In Fig.~\ref{fig:absmag}, in the top panel we show an example of the predicted photometry from our surrogate model (dashed curves) compared to a blackbody SED assumption (assuming the photosphere properties from our surrogates above, square curves) and the true photometry from the {\sc stella} simulations (solid curves). The bottom panel shows $90\%$ credible intervals for LSST u and y bands, demonstrating the range of absolute magnitudes expected for Type II SNe given our trained model. These predictions can enable more optimized follow-up strategies by: (1) predicting optimal observing epochs for maximum signal-to-noise in specific bands, (2) estimating required integration times for spectroscopic follow-up, and (3) helping identify unusual events that deviate from typical Type II behaviour for prioritized observation.
We note that our simulated distributions are also broadly consistent with the absolute magnitudes inferred from ZTF in a volume-limited survey~\citep{Das2025}, which is promising from the perspective of inferences on real observations.
The top panel demonstrates that the discrepancies between our spectra surrogate model and the true inputs are small and likely well captured by the uncertainties on the true simulations as well as those with our surrogate pipeline (as we will discuss later). Furthermore, the discrepancy between the blackbody assumption (square curves) and the true {\sc stella} simulation highlights that the dangers of assuming a blackbody SED, particularly at phases other than from $50-100$ days.

\subsection{Uncertainty estimation}
Our surrogate model architecture enables us to provide an uncertainty on our surrogate model through Monte Carlo dropout~\citeg{Moller2020, Kerzendorf2022}. For example, for our autoencoder model, we can easily generate new spectra for noisy representations of the same latent space vector. Meanwhile for our pure feed-forward neural network surrogates we can use drop out layers to generate different draws from our trained model for the same parameters. This approach can provide us with a time and wavelength dependent uncertainty on our surrogate model. In Fig.~\ref{fig:uncertaintyestimates}, we show an example of the uncertainty on our model predictions on the bolometric luminosity and the spectral luminosity density for the same input parameters. As we see, there is a strong time (frequency) dependence on our uncertainty with the errors in bolometric luminosity largest at early times (as we can also see in Fig.~\ref{fig:residuals}), and at low wavelengths. These uncertainty estimates could be used directly in our inference frameworks to provide robust estimate of parameters marginalising over the uncertainty in the surrogate.

\section{Validation}\label{sec:validation}
We now perform a series of tests on simulated observations (where we know the true input from the simulations themselves) to validate our surrogate models. Here, we will also investigate the implications of the systematic uncertainty in our surrogate models, which was highlighted in Fig.~\ref{fig:residuals} for our luminosity and photosphere property surrogates and discussed above.

Our first validation test involves testing the luminosity, and photosphere temperature and radius surrogate models directly. 
We generate synthetic observations of bolometric luminosity, photosphere temperature, and photosphere radius from the a random sample in the testing data corresponding to a simulation with $M_{\rm ZAMS}=16~M_{\odot}$, ${}^{56}\rm{Ni}=0.08~M_{\odot}$, $\dot{M}=10^{-4.0}$~$M_{\odot}$yr$^{-1}$, $\beta=1.5$, $R_{\rm csm} = 4\times10^{14}$cm, and $E_{\rm sn}=4.5\times10^{51}$erg. Such data could be reconstructed from real observations assuming some bolometric corrections or through multi-epoch SED fitting~\citep[e.g.,][]{Nicholl2018}. Although we do note that such reconstructions are likely to suffer from significant systematics if the assumed SED is incorrect. 
Here, we simply take the true simulation data at random time steps and add random Gaussian noise to the luminosity, temperature, and radius such that each data point has a signal-to-noise ratio of $\approx 5$ and all observations are post $3$ days since explosion (where our residuals become small). Our simulated data alongside the true input from the {\sc stella} simulations are shown in Fig.~\ref{fig:sim_allthree}. 

We perform three independent sets of analyses for this synthetic data. First, we independently fit the luminosity, temperature, and radius assuming a Gaussian likelihood using {\sc Redback}~\citep{Sarin2024} and estimate the posterior using the {\sc pymultinest}~\citep{Feroz2009, Buchner2016} nested sampler through {\sc Bilby}~\citep{Ashton2019}. 
Second, we jointly fit the photosphere properties i.e., the temperature and radius together utilising both surrogates at once with a joint likelihood that is the product of each individual likelihood terms. 
A key aspect we want to test is whether with realistic levels of noise, our systematic error on each surrogate effects our ability to accurately draw inferences on the parameters. A well-performing surrogate model inference framework should produce results consistent with the input independently, with the joint-likelihood analysis providing a tighter constraint, again consistent with the input. 
The joint-likelihood approach also tests another mode of failure with the simple Gaussian likelihood approach: combining two surrogate models the systematic error from both models will stack to create a larger bias, potentially leading to incorrect inferred parameters, we therefore also want to see whether our systematic error on each surrogate can also stack to produce inconsistent posteriors. We note that we only focus on temperature and photosphere jointly here as they are what most commonly will be combined together. 

\begin{figure}
    \centering
    \includegraphics[width=\linewidth]{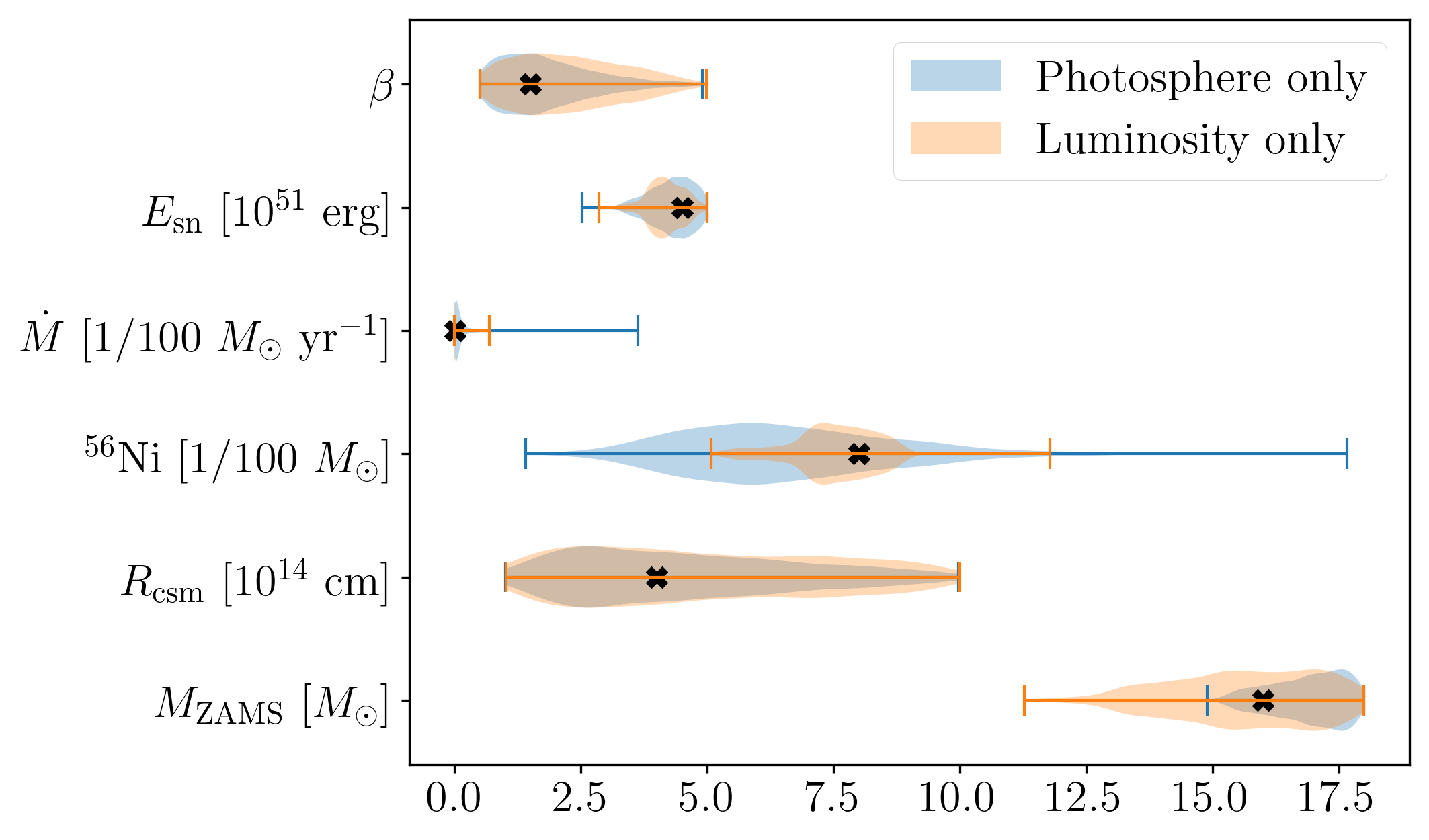}
    \caption{Violin plot showing the one-dimensional marginalised posteriors on a subset of parameters. The blue posteriors indicate results from jointly fitting the photospheric properties, while the red posteriors show results from fitting only the bolometric luminosity. The black crosses indicate the true input, which all our posteriors are consistent with. We note that we have scaled a few parameters to ensure all parameters shown here have a consistent range.}
    \label{fig:simviolin}
\end{figure}

In Fig.~\ref{fig:sim_allthree}, we show the $90\%$ credible interval of the posterior prediction for the bolometric luminosity, photosphere radius, and photosphere temperature independently. These posterior predictions are all consistent with the data and the true input from the simulation highlighting that the surrogate is well-suited to fitting this type of data. We note that each analysis takes $\approx 10$mins. In Fig.~\ref{fig:simviolin}, we show a violin plot of our well-constrained parameters, with the black crosses indicating the true input parameters of the simulation. Again, this consistency highlights that our surrogate model and inference framework functions correctly. The joint likelihood constraints (shown in blue) are also consistent with the input, highlighting that the systematic bias in two surrogates is not significantly affecting our results. 
This study also begins to highlight what we can aim to learn from fitting observations with this model, in particular across a suite of tests, we rarely find $\beta$ to be significantly constrained away from the prior. Moreover, unless the early lightcurve is well-observed, the posteriors on mass-loss rate $\dot{M}$ is also often not informative, meanwhile $R_{\rm csm}$ posterior tends to have a long tail all the way to the edge of the prior. 
We note that a key aspect here is we only fit data post $3$ days since explosion, where the residuals, i.e., the systematic uncertainty (on the luminosity model in particular) are small, including data before this epoch (without accounting for the systematic uncertainty) strongly influences the inferred parameters, and often leads to results inconsistent with the input. 

We now turn our attention to more direct observations: photometry. To be as close to reality as possible, we simulate a Type II SN as it would be observed within the $10$ year LSST survey with Vera Rubin Observatory. Specifically, we take the cadences, filter observing logs, and limiting magnitudes directly from \texttt{rubin\_baseline\_v3.2} survey logs from \texttt{opsimsummary}~\citep{Biswas2020} and use {\sc Redback} to generate synthetic photometry based on these conditions and pointings for a Type II SN at a redshift of $z=0.18$, with $M_{\rm ZAMS}=12~M_{\odot}$, ${}^{56}\rm{Ni}=0.2~M_{\odot}$, $\dot{M}=10^{-3}$~$M_{\odot}$yr$^{-1}$, $\beta=5$, $R_{\rm csm} = 4\times10^{14}$cm, and $E_{\rm sn}=2.5\times10^{51}$erg. 
We fit this data again using {\sc Redback}, however we now do not use our true magnitude errors as the $\sigma$ in the standard Gaussian likelihood, instead, we add an additional error in quadrature to account for the uncertainty in our temperature and radius surrogates. In particular, we now use 
\begin{equation}
\sigma^{2} = \sigma_{i}^2 + \sigma_{\rm sys}^2,
\end{equation}
where $\sigma_{i}$ is the real error on our simulated data and $\sigma_{\rm sys}=0.167$ is the systematic error in the magnitude output of the surrogate model as discussed above. This approach explicitly ignores the time-dependence on the error, but it provides an easy path towards accounting for the uncertainty in our model. We note that we could also put a prior on $\sigma_{\rm sys}$ and marginalise over it if for example we wanted to include additional systematic uncertainty. This approach makes our posteriors more conservative than they otherwise would be, trading precision for accuracy.

\begin{figure}
    \centering
    \includegraphics[width=\linewidth]{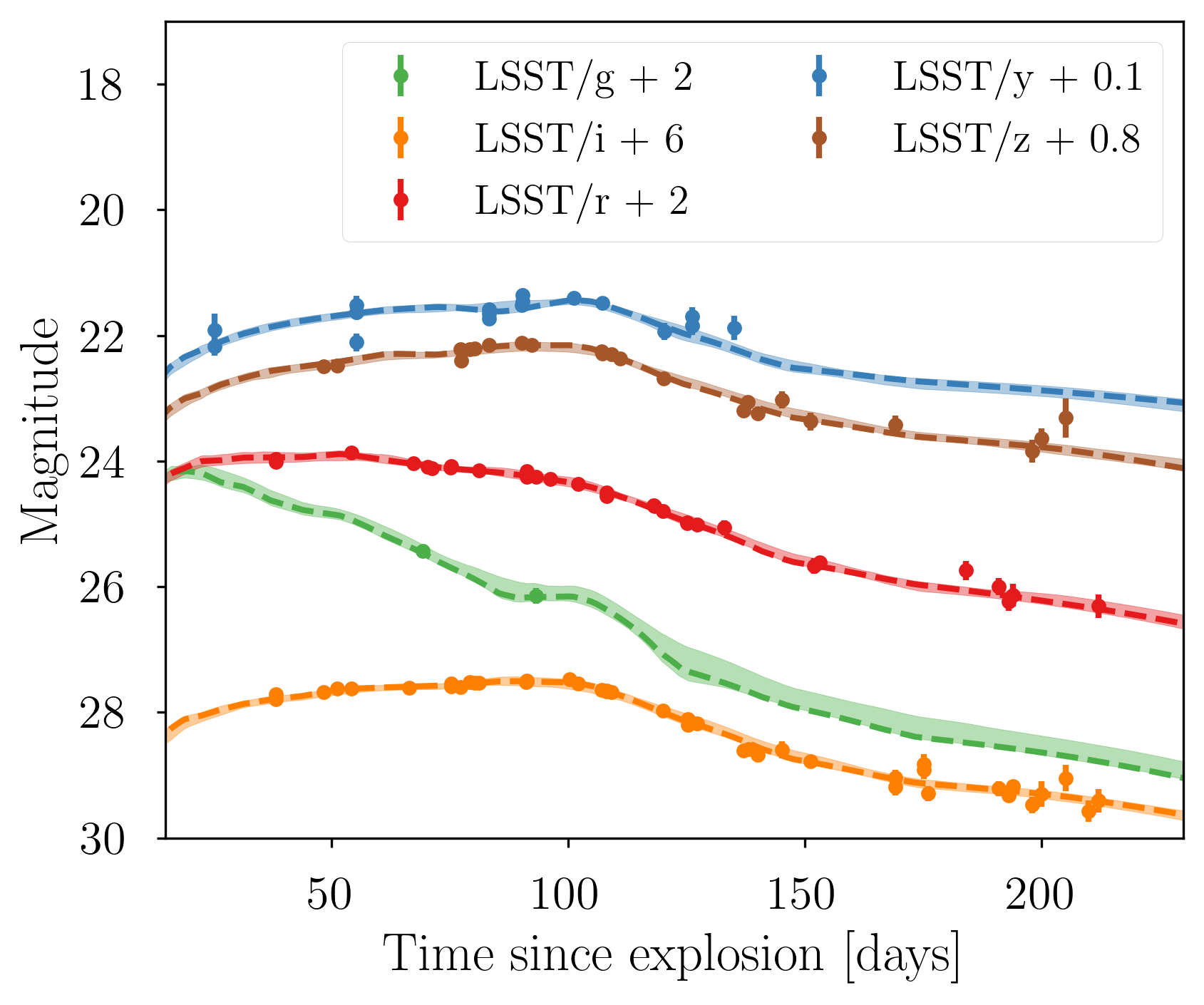}
    \caption{Synthetic Type II SN observed within the 10 year LSST survey with the Vera Rubin observatory. The shaded band indicates the $90\%$ credible interval from our posteriors for each band, while the true input from the {\sc stella} simulations is shown with the dashed curves.}
    \label{fig:lsstim}
\end{figure}

As we discussed earlier, we could provide an estimate of the uncertainty from our surrogate model using Monte Carlo sampling, which we could then use directly in our likelihood instead of a constant $\sigma_{\rm sys}$. However, this approach is slower to evaluate, and we find a constant error approach to provide largely identical results. 
Our simulated data alongside the true input (dashed curves), and the posterior predictions (shaded bands) are all shown in Fig.~\ref{fig:lsstim} for different LSST filters. Our posteriors are consistent with both the data and the true input, highlighting the effectiveness of the model to fit the data and correctly recover the input parameters when considering realistic photometric observations. Furthermore, this study also provides us with some expectations of what constraints we can expect from Type II observations in with the Vera Rubin Observatory. 
Even accounting for our systematic uncertainty on the model, we recover $M_{\rm ZAMS} = 11.88_{-0.45}^{+0.57}~M_{\odot}$, ${}^{56}\rm{Ni}=0.20_{-0.03}^{+0.02}~M_{\odot}$, $R_{\rm csm}=3.84_{-1.07}^{+2.34}~\times 10^{14}\rm{cm}$ and $\log_{10} \dot{M} = -3.14_{-1.44}^{+0.28}$~$M_{\odot}$yr$^{-1}$ ($68\%$ credible interval), respectively.
These constraints are all consistent with the input and represent strong constraints on key parameters. For example, a precision the nickel and progenitor ZAMS mass of $\approx 25\%$ and $\approx 9\%$, respectively. 
We note that the simulated observations above used the proposed \texttt{rubin\_baseline\_v3.2} observing-log from LSST and represents the lightcurve of a typical Type II SNe in LSST with this survey schedule. As such, the precision above may not be representative for all Type II SNe observed by LSST. However, many Type II SNe observed by LSST will likely have follow-up observations with other telescopes and so our constraints could be viewed as a conservative estimate on the expected precision. This paints an exciting picture on the expected constraints from Type II SNe in the future. 

\section{Application to real observations}\label{sec:application}
Having validated our model on synthetic data with known inputs, we now turn to applying our full photometric model to infer the properties of a few well-observed Type II SNe. 
We selected these three well-observed Type II SNe as they have extensive multi-wavelength observations and have been studied by different groups in the past, allowing us to compare our constraints. These events also probe different parts of the parameter space with regards to explosion energy, which is a critical parameter in the model. Moreover, two of these events have strong constraints on the progenitor mass from pre-explosion imaging, which we can directly compare our constraint to.

\begin{figure*}
    \centering
    \includegraphics[width=0.95\textwidth]{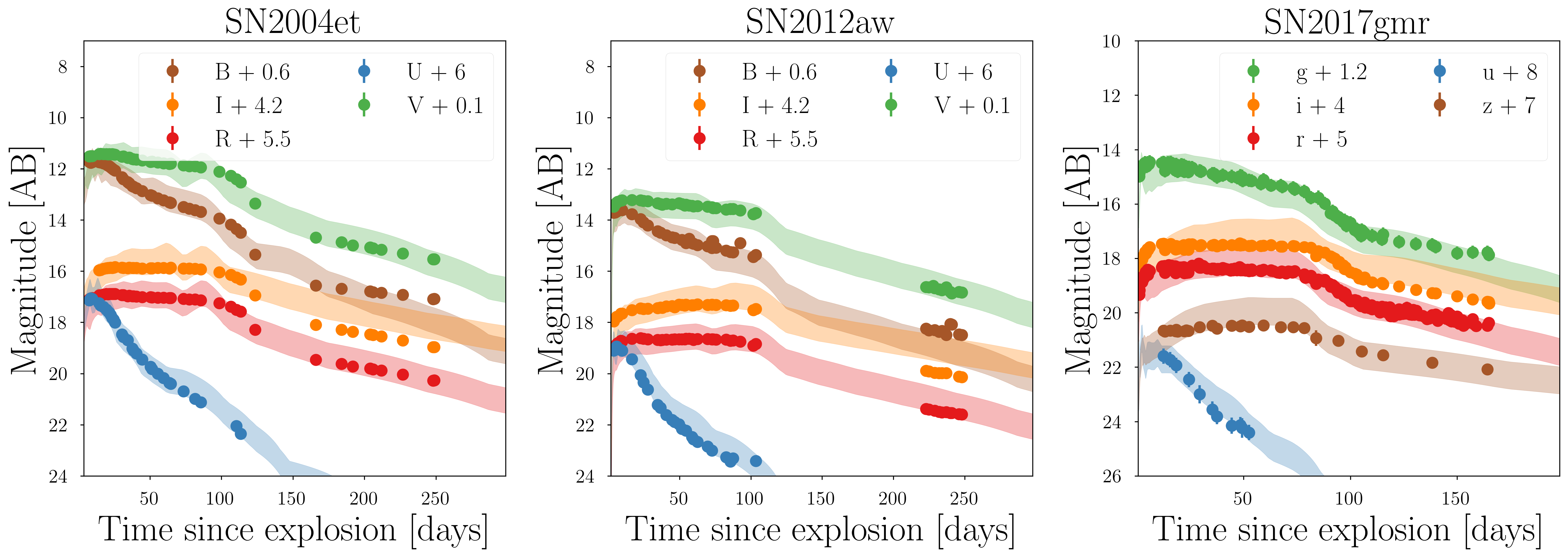}
    \caption{Type II SNe photometry for SN~2004et \citep{Sahu2006}, SN~2012aw \citep{Dessart2013}, and SN~2017gmr \citep{Andrews2019} in the left, middle, and right panels, respectively (data shown in dots for different filters). In the shaded band we show the the maximum likelihood estimate realisation from our fits with the associated $3\sigma$ systematic uncertainty.}
    \label{fig:reallcs}
\end{figure*}

\subsection{SN~2004et}
SN~2004et was discovered in September 2004 and classified as a Type II SN~\citep{Zwitter2004}, at a distance of $\approx 5.5$Mpc~\citep{Li2005} (although this distance has since been updated to 7.72\,$\pm$\,0.10 Mpc~\citep{Tinyanont2019}). Pre-explosion imaging from before the discovery provides strong constraints on the progenitor of SN~2004et of $M_{\rm ZAMS} = 15 \pm 2~M_{\odot}$~\citep{Li2005}. although there are also lower progenitor mass estimates from such observations of $\approx 8~M_{\odot}$~\citep{Crockett2011}.
A number of constraints on the progenitor mass and explosion properties have been placed through spectroscopic~\citep{Jerkstrand2012} and hydrodynamical modelling~\citep{Sahu2006, Utrobin2009, Maguire2010} of the SN, with the latter always providing a high-mass estimate of $\geq 20~M_{\odot}$ inconsistent with any constraints on the progenitor. Hydrodynamical models also place a constraint on the explosion properties and nickel masses with a ${}^{56}$Ni mass estimate of $0.06 \pm 0.04 M_{\odot}$~\citep{Maguire2010} and explosion energies of $1.2 \pm 0.3 \times 10^{51}$erg. 

We model the multi-band photometry up to 250 days\footnote{We assumed a distance of 7.72\,$\pm$\,0.10 Mpc from \citet{Tinyanont2019} and a reddening of $E(B-V)$\,=\,0.41\,$\pm$\,0.05 mag from \citet{Sahu2006}.} of SN~2004et from \citet{Sahu2006} with our surrogate model following the same procedure as we do for our synthetic LSST validation. Our lightcurve fits are shown in Fig.~\ref{fig:reallcs} which highlights a good fit to the data. Our posteriors on key parameters are shown in Fig.~\ref{fig:realviolin}. A noteworthy result of our analysis is the consistency of our progenitor mass estimate with the pre-SN images~\citep{Li2005} and the spectral modelling~\citep{Jerkstrand2012}. In particular, we infer $M_{\rm ZAMS} = 12.15_{-1.06}^{+1.03} M_{\odot}$ which is in contrast to all results based on hydrodynamical models. 
Our constraints on other parameters are all broadly consistent with past results based on hydrodynamical models at $1\sigma$. In particular, we infer a ${}^{56}$Ni mass of $0.09 \pm 0.01 M_{\odot}$, and explosion energy of $1.7 \pm 0.2 \times 10^{51}$erg~consistent with previous results \citep{Maguire2010}.
The consistency of our progenitor mass estimate with pre-explosion imaging but disagreement with other hydrodynamical models could be a product of a few factors. One of the obvious candidates may be the progenitors from~\citet{Sukhbold2016}, which are different to the progenitors assumed in any of the previous papers. Another candidate may be the inclusion (and parameterisation) of extensive CSM in our grid which is in contrast to many simulations in the past. A larger sample of constraints with our model compared to constraints from hydrodynamical modelling and pre-explosion imaging will likely reveal the exact cause of the discrepancy.

\subsection{SN~2012aw}
SN~2012aw was discovered in March 2012~\citep{Fagotti2012} and localised to the M95 galaxy at a distance of $\approx 9.9$Mpc~\citep{Bayless2013, Bose2013}. Again, pre-SN images provide a constraint into the progenitor star with inferred estimates of $12.5 \pm 1.5 M_{\odot}$~\citep{Fraser2012, Kochanek2012, Fraser2016}. Modelling of the lightcurve and spectra provide some constraints on the explosion energy of $1-2 \times 10^{51}$erg and a ${}^{56}\rm{Ni} = 0.06\pm0.1$ $M_{\odot}$ ~\citep{Bose2013}. 
Similar to the previous case study, we fit all available photometry from \citet{Bose2013}\footnote{We assumed a distance of 9.9\,$\pm$\,0.10 Mpc and a reddening of $E(B-V)$\,=\,0.07\,$\pm$\,0.01 mag from \citet{Bose2013}.} using {\sc Redback} and our surrogate model.
Our results are shown in Fig.~\ref{fig:reallcs} and \ref{fig:realviolin}, which are broadly in agreement with constraints. Our progenitor mass constraint $10.61_{-0.32}^{+0.37} M_{\odot}$ is also on the lower end compared to pre-explosion constraints but consistent at $1\sigma$ with the literature~\citep{Fraser2012, Kochanek2012, Fraser2016}. However, we do see a departure at late-times from our model posterior prediction on the lightcurve compared to the data, which is perhaps notable as at these epochs the lightcurve should be driven purely by radioactive decay, which is relatively straightforward to model and infer. There is also a discrepancy in our posteriors on the explosion parameters compared to past results. Here, we infer ${}^{56}\rm{Ni} = 0.09\pm0.1$ $M_{\odot}$ and an explosion energy of $0.63_{-0.04}^{+0.05} \times 10^{51}$erg, both inconsistent with the results based on hydrodynamical models~\citep{Bose2013}.
The discrepancies likely stem from a combination of a few factors: 1) the late-time departure from our model suggests missing physics in the nebular phase, where the LTE assumption in \program{stella} breaks down. 2) Our nickel-mass estimate may be biased by the fixed choice of mixing and (3) the lower explosion energy we infer could reflect differences in how CSM interaction is treated. Future improvements should incorporate non-LTE nebular phase modelling, variable mixing parameters to better match 2012aw-like events, and investigate the role of the CSM with isolated modelling of the early-time data.

\subsection{SN~2017gmr}
We also analyse SN~2017gmr, a Type II SN discovered in September 2017 and localized to NGC 988 at a luminosity distance of $19.6 \pm 1.4$Mpc~\citep{Andrews2019}. Notably, SN~2017gmr is much more energetic than typical Type II SNe with estimates from modelling of the lightcurve and spectra suggesting an explosion energy of $\sim 10^{52}$ erg~\citep{Utrobin2021}, with a nickel mass of $0.02-0.13 M_{\odot}$, higher than either of the above SNe and typical estimates of Type II SNe (0.037$\pm$0.005 $M_{\odot}$)\citep{Osmar2021}. The early lightcurve also shows hints of CSM interaction~\citep{Andrews2019}, which is consistent with systematic analyses of early Type II lightcurves~\citep{Forster2018}. Moreover, polarisation measurements strongly hint at an aspherical ejecta profile~\citep{Nagao2019}.

\begin{figure}
    \centering
    \includegraphics[width=\linewidth]{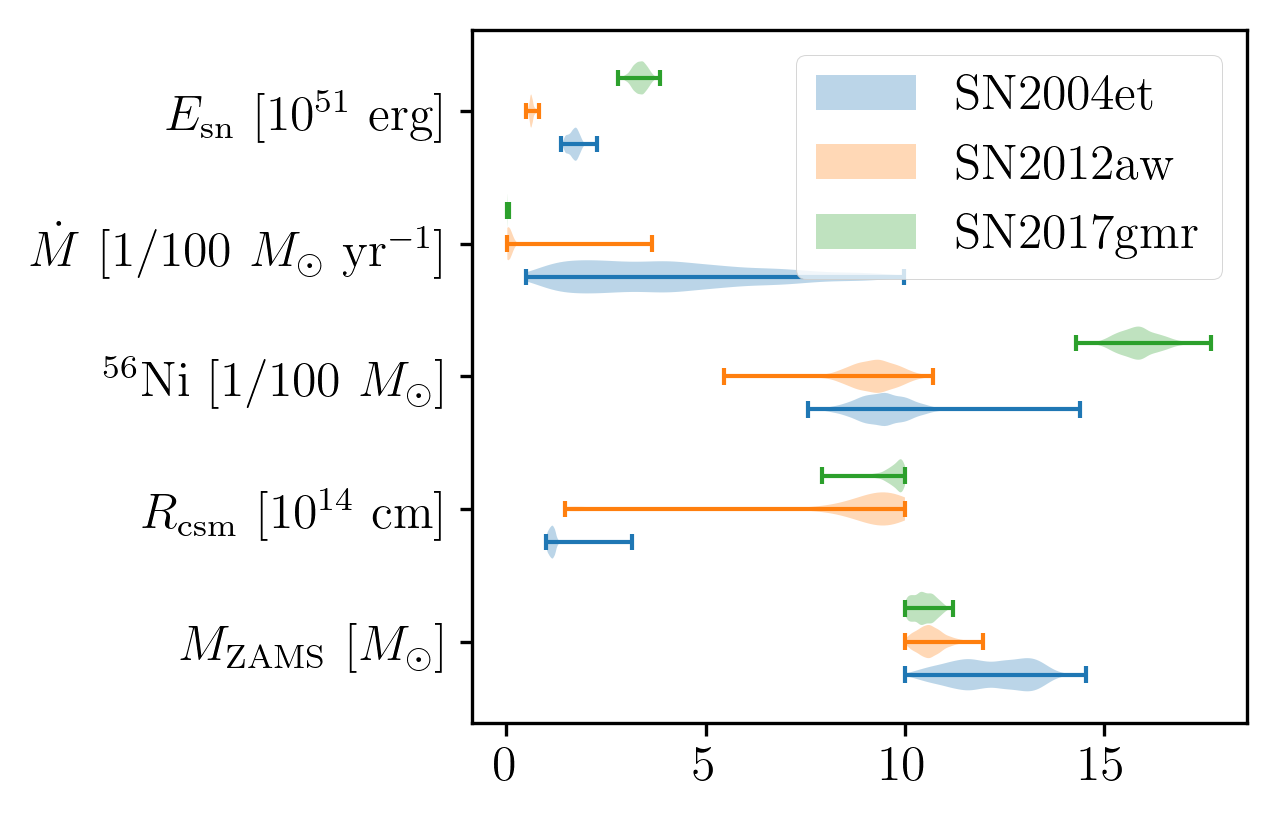}
    \caption{Violin plot showing the one-dimensional marginalised posteriors on a subset of parameters for the three different SNe we looked at in this study. We note that we have scaled a few parameters to ensure all parameters shown here have a consistent range similar to Fig.~\ref{fig:simviolin}.}
    \label{fig:realviolin}
\end{figure}

We fit the light curves\footnote{We assumed a distance of 19.6\,$\pm$\,1.4 Mpc and a reddening of $E(B-V)$\,=\,0.30\,$\pm$\,0.05 mag from \citet{Andrews2019}.} from \citet{Andrews2019} and the results are shown in Fig.~\ref{fig:reallcs} and \ref{fig:realviolin}. Our results are consistent with the idea that SN~2017gmr is more unique and energetic as we infer posteriors on explosion energy of $3.35_{-0.19}^{+0.18} \times 10^{51}$erg and ${}^{56}\rm{Ni} = 0.16 \pm 0.01$ $M_{\odot}$, both higher than the previous two SNe we analysed and broadly consistent with past constraints~\citep{Utrobin2021}. Our nickel mass constraint here is also higher than constraints from hydro models on a sample of Type II SNe~\citeg{Sukhbold2016, Martinez2022}, but consistent with estimates from other methods~\citep{Pejcha2015}. We note that our posterior on $R_{\rm csm}$ strongly rails against the prior at $10^{15}$cm indicating that the model (as currently configured) could provide a better fit (if allowed to extrapolate to higher radii) and correspondingly shift other inferred parameters.

\section{Conclusions}\label{sec:conclusion}

In this work we have introduced and validated surrogate models for Type II SN based on the extensive grid of {\sc stella} simulations from \citet{Moriya2023}. In particular, we have developed distinct surrogate models for bolometric luminosity, photosphere temperature, and photosphere radius, as well as the spectrum (which can be used efficiently to generate photometry). We make these surrogate models and the user-interface to perform inference on observations publicly available via the open-source {\sc Redback} package~\citep{Sarin2024}.

Our different surrogate models offer significant flexibility in how we ultimately analyse observational data. For example, if we have extensive photometric observations across a range of epochs and rest-frame UV to NIR wavelengths and can construct a robust estimate of the bolometric luminosity, then one can use the bolometric luminosity surrogate for parameter estimation. If instead, we have constraints on properties of the photosphere, then we can use the photosphere surrogates or fit different properties with a joint likelihood. Alternatively, and the most likely scenario, if we have photometric observations we can use the spectrum surrogate to directly compare to photometry. 
We note that we could also use the spectrum surrogate for comparisons to spectra. However, our models are not sufficiently high resolution or include important spectral features so this is unlikely to provide a great fit apart from the continuum. 

Through a series of validation tests we have showcased how regardless of the methodology used for inference our different surrogate models all correctly recover the true input (as dictated by the simulations) for realistic observations. In particular, for a synthetic observation of a Type II SN in the 10-year LSST survey, we show how our model can be used to place constraints on the progenitor mass at $\approx 10\%$ precision, even with sparse temporal sampling while marginalising over the uncertainty in our surrogate model. This offers a tantalising prospect to probe the progenitor properties of Type II SN in the near future. Our constraints are limited to the parameters (and by the parameterisation) in the simulations themselves i.e., we can not necessarily constrain properties beyond the parameters in Table~\ref{tab:parameters} or resolve degeneracies in the original simulation, particularly with regards to the parameterisation of the progenitor properties~\citep{Goldberg2019}. 

Applying our model to real observations of SN~2004et, SN~2012aw, and SN~2017gmr, we find that we can generally derive posteriors from the lightcurves consistent with past hydrodynamical models at a fraction of the computational cost. Furthermore, for the specific case of SN~2004et, our constraints on the progenitor masses are consistent with the progenitor mass estimates from pre-explosion images~\citep[e.g.,][]{Fraser2016} in stark contrast to inferences made with any other hydrodynamical model~\citep[e.g.,][]{Sahu2006, Utrobin2009, Maguire2010}.

There are some notable advantages of our design choices for our different surrogate models. For example, the autoencoder could be used to quickly find similar prototypes of new Type II SN with those observed previously to improve classification. For parameter estimation, our design offers several practical advantages. In particular, through Monte-Carlo sampling for the same parameters we are able to efficiently generate an estimate of the error on our surrogate models, including critically, the time-dependence. This error can easily be included in regular inference workflows for example with a Gaussian likelihood, we can add an estimate of the time-dependent $\sigma$ provided by the surrogate in quadrature with the standard error on the data. In practical terms, as the intrinsic models are uncertain themselves we forego this approach for our analyses on real observations above and use a constant model error which provides consistent results but at less computational cost. 

A key computational advantage of our surrogate model is the large capacity for parallelization. In particular, we find that we can generate $\approx 50000$ lightcurves (on a typical laptop) in the same time as it takes to generate one sample. Traditional nested sampling and Markov chain Monte Carlo inference techniques do not take advantage of this large parallelization capabilities and fully leveraging this provides a huge opportunity for inferences at scale. Even ignoring this benefit, we can fit real observations and generate full posteriors in $\approx 15$mins, which is orders of magnitude faster than possible with any hydrodynamical simulation and also faster than some semi-analytical methods~\citeg{Nagy2016}.

Focussing more towards the physics, there are significant improvements we could make in the future. Our general framework for building and validating a surrogate model is agnostic to the input simulations themselves and we could build models for other transients or improve the models here for Type II SNe by replacing the input simulations. In particular, we can retrain our models using a different parameterisation of the progenitor parameters to directly connect with stellar evolution. All simulations that were used in this work assumed ${}^{56}$Ni was mixed into $50\%$ of the ejecta, this will likely vary in reality. Moreover, the degree of mixing is probably correlated with the explosion energy~\citep[e.g.,][]{Eldridge2019, Kozyreva2019}, an effect that newer models should include. An obvious extension is to also include spectral features such that we can compare our surrogate model directly with high-resolution spectroscopy. Furthermore, at late-times as the supernova becomes nebular, the local-thermodynamic equilibrium assumption built into \program{stella} breaks down and instead we could use simulations from software such as \program{sumo}~\citep{Jerkstrand2012} or \program{CMFGEN}~\citep{Hillier2012}, which are more suitable in this regime. This limitation likely affects our constraints on SN2012aw, but is valid limitation for any Type II SN with observations at $\gtrsim 100$ days, where these effects become increasingly important.

Despite the current limitations, the general success of the model and the computational advantage of our surrogate model approach will enable efficient population studies of Type II SNe. Moreover, the general approach introduced here can be readily extended to other transient classes, opening new avenues for constraining stellar evolution and explosion physics in the big-data era.
\section*{Acknowledgments} 
We thank Claes Fransson and Anders Jerkstrand for helpful discussions. We also thank the anonymous referee for their thoughtful comments on the manuscript. N. Sarin and A. Singh acknowledges support from the Knut and Alice Wallenberg Foundation through the ``Gravity Meets Light" project. N. Sarin acknowledges the research environment grant ``Gravitational Radiation and Electromagnetic Astrophysical Transients'' (GREAT) funded by the Swedish Research Council (VR) under Dnr 2016-06012. C. M. B. O. acknowledges support from the Royal Society (grant Nos. DHF-R1-221175 and DHF-ERE-221005).
TJM is supported by the Grants-in-Aid for Scientific Research of the Japan Society for the Promotion of Science (JP24K00682, JP24H01824, JP21H04997, JP24H00002, JP24H00027, JP24K00668) and by the Australian Research Council (ARC) through the ARC's Discovery Projects funding scheme (project DP240101786). Numerical computations were in part carried out on PC cluster at the Center for Computational Astrophysics, National Astronomical Observatory of Japan. 


\section*{Data Availability}
The data for the real observations is compiled from multiple papers referenced in the manuscript and available upon request. The surrogate models are implemented in {\sc Redback}, which is publicly available at \url{https://github.com/nikhil-sarin/redback} and \url{https://github.com/nikhil-sarin/redback_surrogates}. Routines to generate the simulated data used in this paper are also available via this package. We utilised {\sc numpy}~\citep{Harris2020} and {\sc matplotlib}~\citep{Hunter2007} for data analysis and plotting. We used {\sc tensorflow}~\citep{Abadi2016} and {\sc scikit-learn}~\citep{Pedregosa2011} for building our surrogate models. 

\bibliographystyle{mnras} 
\bibliography{paper}
\bsp	
\label{lastpage}
\end{document}